# Teaching the Einsteinian gravity paradigm

Tejinder Kaur[1], David Blair[1], Ron Burman[1], Graeme Gower[2], Elaine Horne[1], Douglas Mitchell[3], John Moschilla[1], Warren Stannard[1], David Treagust[4], Grady Venville[1] and Marjan Zadnik[1]

*[1]The University of Western Australia, Australia*
*[2]Edith Cowan University, Australia*
*[3]The Polly Farmer Foundation, Australia*
*[4]Curtin University, Australia*

*Email: 21358632@student.uwa.edu.au*

# Teaching the Einsteinian gravity paradigm

## Abstract

While Newtonian gravity is an adequate model for current geophysical exploration, Einsteinian gravity, based on the connection between free fall and warped time, has superseded Newtonian gravity as our best understanding of the universe. Einsteinian gravity is fundamental to GPS navigation and is a useful tool for geodesy. The Einstein-First Project is pioneering new curriculum material that seeks to teach students, from ages 11 upwards, the Einsteinian paradigm for gravity. By developing models, analogies and classroom activity based learning, we have found that students are fascinated and easily cope with concepts that adults, indoctrinated with Euclidean-Newtonian concepts, find difficult and confusing. This paper reviews the Einstein-First program, its methods and results of studies with students. We show that the majority of students demonstrate improved conceptual understanding and improved attitude to physics and that female students who enter the program with lower scores than male students, increase their performance to be level with the male students.

## Introduction

Gravitational waves from coalescing black holes were directly detected by huge laser interferometer detectors during 2015 (Abbott et al., 2016). This momentous discovery is the ultimate proof of the century-old Einsteinian model of gravity, which supersedes Newtonian concepts. The year 2016 marks the birth of gravitational wave astronomy (Coleman, 2016). Suddenly, humanity has the technology to listen to the gravitational wave "sounds" of the universe. As the technology improves, we will be able to listen to the entire universe, in space and in time, back to the moment of creation in the big bang.

The question arises: Should the conceptual understanding of the Einsteinian universe be understandable by PhD physicists alone while the majority continue to learn the Newtonian paradigm? The answer is self-evident. The public fascination with the recent discovery shows a demand for understanding. The new signals can be found in the Olympic games, the New York Subway and the Eurovision song contest! The Einsteinian paradigm is our best description of gravity. It could be argued that the Newtonian paradigm is 'good enough' so why create confusion by introducing difficult Einsteinian concepts. Equally an architect could argue that the flat Earth paradigm is good enough so why bother teaching that the Earth is spherical. The answer again is self-evident. We have an obligation to teach our best understanding of the universe.

For several years a group of educationalists in Western Australia have been developing an "Einstein-First" curriculum for schools. Our goal is to reverse the current historical approach to teaching science. We aim for students to learn the fundamental Einsteinian concepts from the age at which they begin to conceptualise space, time, light and matter. We found that students are entirely receptive to Einsteinian concepts (Kaur, in prep.). The key concepts: curved space, warped time, photons and quantum behaviour are unsurprising to younger students (Year 6). However for older students (Year 11) and even more for adults trained in the worldview of flat space, absolute time and mechanistic interactions the new paradigm is surprising. Students assert that they are not too young to learn these things, and these topics should be in the curriculum (Pitts et al., 2014).

In the following, we first give a simple argument for the connection between gravity and warped time, and then summarise a methodology for introducing Einsteinian ideas to young people. The next section briefly describes the Einstein-First program followed by a review of results from trials conducted with students from ages 11 to 17.

## Gravity as Warped time

In the Einsteinian paradigm, gravity is not a force that emerges out of planets but a consequence of the warping of time created by masses. When we climb a staircase of height h, we increase our potential energy by mgh, but this arises not from working against a gravitational force but from working against a time gradient. This is easy to demonstrate by combining Einstein's famous mass-energy equation with Schwarzschild's solution for the space-time metric near the surface of the Earth.

The change in time for height h above the Earth surface (for small height differences) is given by

$$\frac{\Delta t}{t} = \frac{gh}{c^2}$$

while, the energy of a mass m is given by:

$$E = mc^2$$

Combining these results we see that Newtonian potential energy mgh is the product of the mass energy and the time dilation factor:

$$\frac{\Delta t}{t}.mc^2 = mgh$$

This equation shows a link between gravity and time dilation. Gravitational potential energy can be understood as a direct manifestation of time dilation.

The time differences associated with height are now a part of every GPS navigation system (Neil, 2003) and modern atomic clocks are being designed for use in geodesy (Cacciapuoti and Salomon, 2011). Thus, Einsteinian gravity is already a part of modern geophysics.

In the Einstein-First program we introduce students step-by-step to the concepts that led to the concept of gravity as time dilation. We do this using many models and analogies as discussed in the next section. These models and analogies first introduce students to the *idea* of curved space, *measurements* of curved space and then we go on to introduce warped time in both real and thought experiments with clocks.

Once students have learnt the best modern understanding of gravity, it is natural to introduce the useful approximations that will be important tools in most areas of science and engineering. These are the approximations of Euclidean geometry and Newton's law of gravitation. Both of these approximations are extremely accurate on Earth but their breakdown is easily observed by modern astronomers and they are seriously inaccurate near a black hole.

**The Einstein-First program**

The Einstein-First program has focussed particularly on developing simple, inexpensive activities that use analogies and models to teach the concepts of Einsteinian physics. To teach the fundamental concept of general relativity – *matter tells space-time how to curve; space-time tells matter how to move* (Wheeler, 2000), we use large stretched lycra sheet and golf balls to show how mass produces curvature in the fabric of the sheet, which is a 2-dimensional analogy for space-time. Figure 1 shows students working with a large lycra sheet. This model is the basis for many experimental investigations.

Using the lycra sheet model, students can observe that curvature increases with an increase in mass. They can learn about circular and elliptical orbits, three-body interaction and even test the inverse square law of gravitation. By using toy cars with fixed steering to represent light beams, they can see how the paths of cars are deflected in curved space. This is compared with easily accessible images of gravitational lensing by the Hubble space telescope.

When using models it is important to point out their limitations. In the lycra sheet model, the apparent gravitational forces are created through the components of the gravitational force acting on each ball. The warping of time, essential to understanding gravity on Earth, is missing.

It is impossible to introduce Einsteinian gravity without students becoming confident with ideas of geometry in curved space. Having introduced the idea that matter curves space, we now undertake more quantitative activities to investigate geometry on curved space. To do this we use Chinese cooking woks as models for 2-dimensional curved space. Balloons can also be used for such experiments but are more difficult to use for quantitative experiments.

Steel woks, combined with magnetic posts taken from magnetic construction toys, are particularly useful for allowing students to investigate the concept of a straight line in curved space. Figure 2 shows students surveying straight lines on the curved space and verifying that these are the shortest paths using a stretched string. By constructing a triangle with three straight lines, students measure the perimeter using a piece of string and the angles with a protractor.

We ask students to plot a graph of the sum of the angles of various triangles as a function of the perimeter. A typical set of data is plotted in Figure 3. Students observe that small triangles converge to those that they would draw on flat paper (Euclidean geometry) and in general as the perimeter increases, the sum of the angles of a triangle increases from 180 degrees to almost 360 degrees. Through these activities, students discover that Euclidean laws of geometry are a special case. They obtain an intuitive understanding of geometry of the universe through experimental geometry, and this prepares them for formal studies of Euclidean geometry. They learn that Euclidean geometry is merely an excellent approximation near the Earth.

At the end of our curved space activities, we relate to them Karl Gauss's pioneering experiment of 1827, in which he tried to measure the shape of space by using sunbeams to construct a triangle between the peaks of three mountains (Gauss, 1827). This enables students to make the connection between the real world and the two dimensional spaces they worked with, and to visualise the idea of three-dimensional curved space. Once students have obtained experience with woks we refer them back to the lycra sheet experiments where the toy cars mapped the paths of photons.

The most difficult step in bringing students to an understanding of Einsteinian gravity is to show how time dilation is the cause of the fictitious gravitational force. We first introduce students the equivalence principle – the fact that all objects fall at a speed independent of their weight.

This was recognised by Galileo in the context of dropping objects from the Leaning Tower of Pisa. At the time of Galileo, most people believed Aristotle's statement that things fall with constant speed proportional to their mass. Students can test this statement by dropping two objects with a large mass difference. In the classroom, students do an experiment by dropping empty and full water bottes. In our project we are fortunate to have access to the Leaning Tower of Gingin at the Gravity Discovery Centre where students can do real Galileo-scale experiments using balloons full of water, and study other effects such as wind resistance and crater formation.

Newton explained gravity as a force. Einstein pointed out that gravity and acceleration were indistinguishable. In his theory, gravity is the force that has to be applied to prevent free fall, while mass is the cause of free fall trajectories. These trajectories arise, not because of an external force, but because the time coordinate is warped.

We introduce warped time by discussing its experimental observation by spacecraft such as Gravity Probe A and the GPS satellites. Then we use thought experiments with perfect atomic clocks to discuss time differences at different heights above the Earth. We introduce the Einsteinian concept of gravity using Einstein's "first law of motion", which states that all free fall trajectories are the shortest path in space-time. This can be formulated as the principle of maximal aging, which states that freely falling objects take the path of maximal aging. These considerations, combined with the knowledge of gravitational time dilation, gives recognition that a map of space-time is distorted, as shown in Figure 4. Because the map is distorted, curved paths become the shortest distance just as the shortest distances between points on planet Earth (such as aircraft trajectories) look curved when presented on a flat space map projection. This point has been extensively explored recently by Gould (2016). To develop the above ideas we begin with activities in which students plot distance-time maps for simple journeys (with one dimension of space only), then consider the arbitrariness of our units, and how we must connect our time units with our distance units. Students deduce the need for a speed to make this connection, and deduce that the speed of light provides a universal connection between space and time.

In addition to the above models and analogies for teaching gravity, we have also developed models and analogies to teach quantum physics, which we will not discuss in this paper.

The Einstein-First program has been trialled across ages from Year 6 to Year 11. A 10-week program was run with Year 9 students in two different years at School 1, Western Australia. A whole day program was delivered to National Youth Science Forum students in 2013 and 2014 and a one-week program was delivered at School 2. All classes have used small group activities, working with models and analogies. Students were encouraged to participate in group discussions. Three professional development programs were run with both general science and physics teachers. Altogether, more than 400 students and 60 teachers have participated in our programs.

The purpose of the above trials was to research students' ability to understand Einsteinian physics and to assess student attitudes towards the program. The teachers programs were designed to assess teacher ability to accept and deliver such programs. To measure the effectiveness of the Einstein-First program, we assessed students' ability to comprehend the key concepts through conceptual understanding questionnaires and their attitudes to physics and to specific topics using attitude questionnaires. In the conceptual understanding questionnaire, we asked simple questions such as "What is gravity?", "Can parallel lines ever meet?". While in the attitude questionnaire we asked questions such as physics is a difficult subject for me and I think physics is an interesting subject. Some of the results from the trials are presented in the next section.

**Results**

Here we will summarise some results from the Einstein-First trials with Year 6, Year 9, Year 10 and Year 11 students. Three different teachers ran programs with different years of students.

Firstly, we will present some attitude results for studies with Year 6 students who attended six lessons on Einsteinian concepts. The results reported here were taken from data given in Pitts et al. (2014). The attitude questions asked by the authors were "Was it interesting to find out about space and time and gravity?" and "Do you feel you are too young to understand Einstein's ideas?" Figure 5 and Figure 6 present results obtained from these two questions.

Figure 5 shows that majority of students responded that the concepts of space, time and gravity are very interesting. Only one student identified the program as boring. Figure 6 shows responses to the "too young" question. Five members of the class thought that they were too young to understand Einsteinian physics; the majority of the class (14) thought that they were not too young to understand Einsteinian physics and others were undecided. Pitts et al. concluded that Year 6 students were potentially able to comprehend Einsteinian concepts.

Next we present the results on Year 9 students conceptual understanding based on pre and post-questionnaires. These students were academically talented and attended a 10-week program on Einsteinian physics in 2014. Figure 7 shows student understanding of Einsteinian concepts before and after the program. It is clear from the Figure that students' conceptual understanding improved dramatically after the program. Student pre-test scores were very low, with less than 10% of students scoring more than 50%. It is interesting that some of the students who had the lowest initial scores achieved results as high as the students with the highest pre-test scores. After the program, 53 out 57 students scored 80% or above.

We found that boys had slightly greater interest in physics throughout the program, but that girls showed a significantly greater *increase* in interest. Figure 8 gives the data showing that girls' interest toward physics improved from 50% to 80%, much greater than the boys. No evidence was found that the material was too complex or that students were too young for these topics. The classroom teachers present during the program were highly motivated by activities used in the program.

We undertook a shorter program with Year 10 academically talented students at School 1. This program was focussed on the Einsteinian physics behind the production of gold in the universe, and its special properties which can only be explained using relativity. In this case, students' initial scores were in the range 15-50%. Again, the students in the class improved significantly after the program. As shown in Figure 10, the improvement by girls again exceeded that of the boys. In this case, the girls performed lowest in the pre-test but exceeded the boys in the post-test.

The final case we discuss is a study of one-day program with Year 11 students who were participants of the National Youth Science Forum. They were advanced students and selected from different states of Australia. Figure 11 presents results from conceptual understanding questionnaires. The results in Figure 11 show a similar trend to the results from younger students. Initial conceptual understanding was low, although a greater fraction of the class scored near to 50%. The post-test results show high final scores. However, the scatter of results probably implies that a single day program is insufficient for consolidating new ideas.

Overall, we found that most of the students improved their conceptual understanding after the program and students demonstrated their interest towards learning Einsteinian physics.

## Conclusion

The long-awaited direct detection of gravitational waves, with its discovery of the first black-hole binary merger, provides new incentive to teaching Einsteinian physics to all students. Gravitational waves are waves of gravity gradient, and cannot be explained by Newtonian physics. The Einstein-First program has shown that it is possible to present the concepts of Einsteinian gravity to school students from ages 11 to 16. Now that Einsteinian gravity has replaced the Newtonian description of gravity, we believe that we have created a set of tools to allow us to teach young people our best understanding of the universe. Our trial programs have shown that the majority of 11-year-old students did not consider themselves too young to learn the fundamental concepts, while 15 year olds demonstrated impressive conceptual understanding after a 10-week program. High school female students entered the programs with lower performance than the male students but universally showed a greater improvement than male students. This is evidence that modernising the school science curriculum will have significant benefits for gender equity in physical science.

## Acknowledgements

We wish to thank the teachers from our partner schools for facilitating the classroom trials that made this work possible. The research was funded by the Australian Research Council and supported by the Gravity Discovery Centre Foundation and the Graham (Polly) Farmer Foundation.